\documentclass[prl,twocolumn,showpacs]{revtex4}
\usepackage{amsmath}
\usepackage{epsfig}

\def\b{\bar}
\def\d{\partial}

\def\cD{{\cal D}}

\def\m{\mu}
\def\n{\nu}

\def\t{\tau}

\def\~{\widetilde}

\def\bY3{\bar Y_{,3}}
\def\Y3{Y_{,3}}
\def\z{\zeta}
\def\Z{{\b\zeta}}
\def\Y{{\bar Y}}
\def\cZ{{\bar Z}}
\def\be{\begin{equation}}
\def\ee{\end{equation}}
\def\bea{\begin{eqnarray}}
\def\eea{\end{eqnarray}}

\def\fn{\footnote}

\def\cF{{\cal F}}

\def\mn{{\mu\nu}}

\begin{document}

\title{Aligned electromagnetic excitations of a black hole \\
 and their impact on its quantum horizon}

\author{Alexander Burinskii$^{1}$,
Emilio Elizalde$^{2}$, Sergi R. Hildebrandt$^{3}$, Giulio
Magli$^{4,}$\footnote{E-mail addresses: bur@ibrae.ac.ru,
elizalde@ieec.uab.es, srh@iac.es, magli@mate.polimi.it}}

\affiliation{$^1$Gravity Research Group, NSI, Russian Academy of Sciences, B. Tulskaya 52  Moscow 115191 Russia\\
$^2$Instituto de Ciencias del Espacio (CSIC) \& Institut
d'Estudis Espacials de Catalunya (IEEC/CSIC)\\ Campus UAB, Facultat de Ci\`encies, Torre C5-Parell-2a planta, 08193 Bellaterra
(Barcelona) Spain\\
$^3$Instituto de Astrof\'\i sica de Canarias,
C/V\'\i a L\'actea s/n, La Laguna, Tenerife, 38200, Spain\\
$^4$Dipartimento di Matematica del Politecnico di Milano,
Piazza Leonardo Da Vinci 32, 20133 Milano, Italy
}

\begin{abstract}
We show that elementary aligned electromagnetic excitations of
black holes, as coming from exact Kerr-Schild solutions, represent
light-like beam pulses  which have a very strong back reaction on
the metric and change the topology of the horizon. Based on York's
proposal, that elementary deformations of the BH horizon are related
with elementary vacuum fluctuations, we analyze deformation of the
horizon caused by the beam-like vacuum fluctuations and obtain a
very specific feature of the topological deformations of the
horizon. In particular, we show how the beams pierce the horizon,
forming a multitude of micro holes in it. A conjecture is taken
into consideration, that these specific excitations are connected
with the conformal-analytic properties of the Kerr geometry and
are at the base of the emission mechanism.
\end{abstract}

\maketitle

\section{Introduction} The interest in Hawking radiation from a
black hole \cite{Haw,Sol,RobWil,Sta,JWC,Rov,ZZ} is not decreasing.
New derivations of this law, some of them very different from the
original one and opening new avenues in quantum physics, have
appeared in the last years. Even more interesting, from the
experimental (i.e., observational) side results are quickly
reaching limits which clearly indicate that some specific
properties of black holes will be testable in the near future.

There are a few semiclassical explanations of the origin of the
quantum evaporation process of BHs. Some of the most popular are:
(i) Evaporation is related with `quasi-normal' modes of black hole
excitations \cite{deWit,BirDev} and the emitted particles are
radiated by means of a quantum tunnelling process from under the
horizon \cite{BirDev}. (ii)  The loop quantum gravity approach
\cite{ABK,ABCK} proclaims the quantum horizon, experiencing
topological fluctuations, to be itself the source of the quantum
radiation. (iii) Strong fields near the horizon lead to virtual pair
creation in its outer vicinity, and while one of the particles falls
into the horizon the other keeps going to infinity
\cite{BirDev,DerRuf,Gib}. This process can also be considered as a
quantum tunneling via a classically energetically prohibited region,
the Klein paradox \cite{DerRuf}.

In the original derivation of BH evaporation, Hawking described
thermal radiation as quantum tunneling triggered by vacuum
fluctuations, however, the treatment was based upon a fixed
background, without a back reaction causing the fluctuations of the
background geometry. Consideration of the back reaction of vacuum
fluctuations and their impact on the horizon was initiated by York
\cite{Yor} who used spherical zero-point fluctuations (ringing
modes) and the corresponding deformation of the space-time metric.
This dynamical approach was also supported by loop quantum gravity
\cite{Rov} and in subsequent work \cite{ZZ,JWC}(see also the
references in \cite{JWC}). However, the treatment was restricted
to simplest deformations of the horizon, caused by spherical or
ellipsoidal quasi-normal modes. Meanwhile, recent analysis of
rotating black holes \cite{BEHM} in the Kerr-Schild formalism
showed that the {\it exact} electromagnetic excitations on the
Kerr background has to be {\it aligned with the Kerr congruence.}
As a consequence, the elementary electromagnetic excitations turn
out to be very far from the usual spherical harmonics. They
constitute a brand new basis for elementary excitations and for
elementary deformations of the horizon.

The present treatment,  based on these exact solutions,
shows that horizons acquire a very specific topological
deformation subjected to influence of  vacuum electromagnetic
excitations. We show that an elementary aligned excitation is
accompanied by the formation of a singular light-like beam, which
emerges from the black hole. The influence of those beams on the
horizon was analyzed in our previous paper \cite{BEHM}. It was
shown there that the horizon is very sensitive to such beams,
which form thereby narrow holes connecting the inner and
outer regions. Even weak elementary beams may be able to break
through the Kerr black hole horizon. Moreover, the elementary
deformations of the horizon turn out to be of topological nature.
We consider also the wave aligned electromagnetic excitations of
the Kerr solution and show that they are asymptotically exact in
the low-frequency limit. This result allows us to use aligned wave
excitations for modeling excitations of massive black holes
through weak fields of virtual photons, and to arrive to a new
feature of deformation of the horizon, by invoking a large set of
microscopic holes in the horizon, caused by singular beams.

We put forward the conjecture that the complex-analytic structure of
the Kerr congruence and of the wave excitations which are aligned
with the Kerr congruence are at the base of the emission mechanism.

\section{Aligned electromagnetic solutions in the Kerr-Schild
geometry} As is well known \cite{DKS,BEHM,Multiks}, the Kerr
metric is the unique regular one representing a much broader
Kerr-Schild class of Einstein-Maxwell solutions. There is an
infinite set of exact rotating Kerr-Schild solutions
containing semi-infinite light-like beams (the singular
$pp-$waves, see e.g. \cite{BurAxi}). The electromagnetic field of
these solutions is aligned  with the principal null congruence of
the Kerr geometry. We show in this letter that correspondingly
aligned wave excitations of the Kerr geometry tend to the above
mentioned exact stationary Kerr-Schild solutions in the
low-frequency limit. Below, we summarize the discussion and
specify the whole argument for the case of the Kerr-Newman
solution in the Kerr-Schild  form.

The peculiarities of the Kerr-Schild approach rely on the
Kerr-Schild {\it Ansatz} for the metric, which is adjusted to an
auxiliary Minkowski space-time with a metric $\eta_\mn ,$ contrary
to other approaches, in which the coordinate system is adjusted to
a position of the horizon which is unstable itself and, as we will show,
is very sensitive to the electromagnetic field. The Kerr-Schild
approach allows one to see some peculiarities of the solutions
which are not seen in other approaches. In particular, it exhibits
complex-analytic properties of the black hole geometries which are
related with two sheets of the space-time. On the Kerr-Schild
background one sees that the Kerr geometry is two-sheeted, having
an `in' $(-)$ and an `out' $(+)$ sheet, and corresponding $
g^{+}(x)$ and $ g^{-}(x)$ metrics: \be g^{(\pm)}_\mn =\eta_\mn -2H
k^{(\pm)}_\m k^{(\pm)}_\n, \label{KSpm}\ee which are different on
the `in' and on the `out' sheets. Moreover, the fields on the `in'
and `out' sheets do not interact with each other. In particular
the principal null directions $k^{(\pm)}_\n (x)$ are different on
different sheets, being determined by different functions $Y^+(x)$
and $Y^-(x),$ in the null Cartesian coordinates $u=z-t,\quad
v=z+t,\quad\zeta=x+iy,\quad\bar\zeta=x-iy,$ namely \be
k_\m^{(\pm)} dx^\m = P^{-1}(du +\Y^\pm d\z + Y^\pm d\Z - Y^\pm
\Y^{(\pm)} dv). \label{kpm}\ee The known Kerr singular ring is a
branch line of space, and forms a gate to the negative sheet of
the metric. The Kerr principal null congruence (PNC) propagates
from the `in' sheet of the Kerr space to the `out' sheet through
the Kerr singular ring, covering
the auxiliary Minkowski space-time twice. This peculiarity of
space-time has not been observed earlier by analysis of the usual
spherical black hole solutions and leads to far-reaching
consequences, as we intend to show.

The electromagnetic field corresponding to the exact Kerr-Schild
solution has to be {\it aligned} with the Kerr principal null
congruence. This means that the Maxwell tensor of strength
$F_\m^\n$ and vector potential $A_\m$ satisfy the following
alignment condition
 \be k^\m F_\m^\n =\kappa k^\n , \quad k^\m A_\m=0, \label{align}  \ee
where $\kappa$ is some scalar function.
 Now, since the
vector fields $k^\pm _\m$ are different on the `in' and `out'
sheets of the Kerr geometry, one sees that the `in' and the `out'
aligned fields cannot possibly be on the same sheet simultaneously,
in accordance with the structure of the algebraically special
Kerr-Schild solutions. Therefore, in the Kerr geometry the `in'
and the `out' modes of the aligned electromagnetic waves are
positioned on the two different sheets of the real Kerr space and
do not interact (at least classically). The alignment condition is
very strong and leads to important physical consequences. It
resembles the condition for exclusion of the longitudinal modes in
dual-string modes and is the origin of the complex-analytic
properties of the Kerr geometry.\fn{Another source of complex
analyticity is the Kerr theorem which determines the Kerr
congruence, $k^\m(x),$ in terms of complex-analytic surface in
twistor space \cite{KraSte,Multiks}.}

Note that, in the ordinary perturbative approach based on
quasi-normal modes, the important two-sheetedness of the Kerr
geometry has been ignored. The perturbative approach breaks
analyticity of the solutions and leads to a drastic change in the
structure of the usual quasi-normal and aligned electromagnetic
excitations of (rotating) black holes. In particular, the analytic
aligned wave excitations lead to the formation of light-like
singular beams emanating from the black hole and changing the
properties of its horizon, breaking up the usual classical
structure of the black hole.

We show here that the aligned wave solutions are asymptotically
exact for slowly varying excitations, that is, in the
low-frequency limit. Similar beams will also appear for the
aligned excitations of the rotating sources without horizons
considered in \cite{BEHM,BurAxi}.

\section{Exact Kerr-Schild solutions with singular beams} For the readers'
convenience we recall here the structure of the exact Kerr-Schild
solution displaying the aligned electromagnetic field which
produces the singular beams \cite{DKS}. The exact electromagnetic
field which is aligned with the Kerr null congruence, $k^\m$,
depends on an arbitrary holomorphic function $\psi(Y),$ which
determines the Kerr-Schild (KS) metric (\ref{KSpm}) via the
function \be H =\frac {mr - |\psi|^2/2} {r^2+ a^2 \cos^2\theta} .
\label{Hpsi} \ee In the Kerr-Newman solution, $\psi =q$ is the
charge. However, any holomorphic function $\psi(Y) $ yields also
an exact aligned solution. The function \be Y=e^{i\phi} \tan \frac
\theta 2 \ee is a projective angular coordinate. It determines the
Kerr congruence (\ref{kpm}) and also the KS tetrad $e^a$ with the
real directions \be e^3=Pk_\m dx, \quad e^4 =dv + h e^3, \quad
h=HP^{-2}, \label{e34H}\ee
 and the mutually complex conjugated forms
\be e^1 = d \zeta - Y dv, \qquad  e^2 = d \bar\zeta -  \bar Y dv.
\label{e12} \ee
For the Kerr congruence satisfying the {\it geodesic} and {\it
shear-free} conditions: \be Y,_2=Y,_4=0,\ee the Einstein-Maxwell
field equations were integrated
 by Debney, Kerr and Schild \cite{DKS} in a general form.
 They reduced them to a system of two equations for the electromagnetic field
 and two more for the gravitational field.
The electromagnetic field aligned with the Kerr congruence admits
the self-dual tetrad components, \be\cF _{12}=AZ^2, \quad \cF
_{31}=\gamma Z - (AZ),_1, \label{E6}\ee where $\cF_{ab}= e_a^\m
e_b^\n \cF_\mn ,$ and the function $Z$ is a complex expansion of
the congruence, which for the stationary Kerr-Newman solution is
inversely proportional to a complex radial distance  $ Z= -P
/(r+ia\cos \theta),$ where $P= 2^{-1/2}(1+Y\Y).$ Functions $A$ and
$\gamma$ obey the Eqs.~(\ref{3}),(\ref{4}), given in App.~B.
Explicit solutions were obtained in \cite{DKS} for the case
stationary fields without wave excitations,\fn{About this problem
see also \cite{Ker}.} which corresponds to $\gamma=0,$ and \be A=
\psi/P^2 , \quad \psi,_2 =\psi,_4=0 .\label{E3} \ee The function
$\psi(Y)$ is analytic and can be represented as an infinite
Laurent series: $\psi(Y)= \sum _{n=-\infty}^{\infty}q_n Y^n. $

If the function $\psi(Y), \ Y\in S^2 ,$ is not a constant, it has
to contain at least one pole which may also be at $Y=\infty$ (or
$\theta =\pi$). So, except for the Kerr-Newman solution, for which
$\psi(Y)=q=$const, the solutions $\psi(Y)$ need to be singular at
some angular directions $Y_i =e^{i\phi_i}\tan \frac {\theta_i} 2
$. We then consider the {\it Ansatz}: \be \psi (Y) = \sum _i \frac
{q_i} {Y-Y_i}, \quad A=\psi(Y)/P^2. \label{E5}\ee Singular
electromagnetic fields have a strong back reaction on the metric,
via the function $\psi(Y)$ in (\ref{Hpsi}).
 For $q_i=$const, these functions yield exact, self-consistent solutions
of the full system of Kerr-Schild equations.

 In this case, the term $\cF_{31},$ which
describes the null em-radiation $\cF_\mn =\cF_{31}e^3_\m e^1_\n $
along the Kerr congruence,  takes the form
\bea \cF_{31}&=&(AZ),_1 = \sum _i \{-(Z/P)^2 \frac {q_i} {(Y-Y_i)^2} \nonumber \\
&-& 2 (Z/P)^2 \frac {q_i P_Y} {P(Y-Y_i)}  \nonumber \\ &+&  2ia \Y
 \frac {q_i} {Y-Y_i} P^{-2}(Z/P)^3 \}.\eea
Since $Z/P = 1/(r+ia\cos\theta),$ this expression is singular at
the Kerr ring and falls off as $r^{-2}$ and $r^{-3}$ with the distance.
The first term in the sum is leading. It contains the poles $ \sim
\frac {q_i} {(Y-Y_i)^2}$ which are singular along the lines of
congruence in the direction $Y_i $ and do not fall off, taking
asymptotically the form of singular $pp$-wave Peres solutions
\cite{KraSte} describing the narrow singular electromagnetic beams
in each of the angular directions $Y_i$ \cite{BurAxi}, i.e.
semi-infinite `axial' singular lines which destroy the horizon and
which lead to the formation of topological holes in it.

 The properties of the horizons of these solutions
were considered in \cite{BEHM}. It was  shown that the black hole
horizon is pierced by the axial singularity, with the appearance
of a tube-like region which connects the internal with the external
region, thus possibly allowing matter to escape from the BH. As a
result, the Kerr singularity turns out to be ``half-dressed''. The
structure of the horizons  for the solutions containing axial
singular lines follows from the Kerr-Schild form of the metric for
the function $H= \{mr-\psi(Y)^2/2\}/(r^2 + a^2 \cos^2 \theta)$,
where the oblate coordinates $r,\theta$ are used on the flat
Minkowski background $\eta^{\mn}$. For the rotating solutions, the
horizon splits into four surfaces. Two of them correspond to the
statical limit, $ r_{s+}$ and $r_{s-}$, determined by the
condition $g_{00} =0$, and the other two surfaces correspond to
the event and Cauchy horizons, $S(x^\m)=$ const, which are the
null surfaces determined by the condition $ g^\mn (\partial _\m S)
(\partial _\n S) =0. $

The simplest axial singularity is the pole $\psi =q/Y.$ In this
case, the boundaries of the ergosphere, $r_{s+}$ and $r_{s-}$, are
determined by  $g_{00} =0$  and the solution acquires a new feature:
the surfaces $r_{s+}$ and $r_{s-}$ are joined by a tube, conforming
a simply connected surface. The surfaces of the {\it event horizons}
are null ones and obey the differential equation $(\d_r S)^2 \{ r^2
+a^2 +(q/\tan \frac {\theta} {2})^2 -2Mr \} - (\d_{\theta} S)^2 =0.$
The resulting structure for the horizon is illustrated in Fig.~1
(taken from \cite{BEHM}).

\begin{figure}[ht]
\centerline{\epsfig{figure=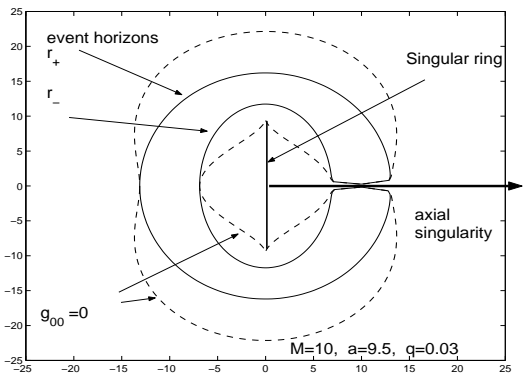,height=5cm,width=7cm}}
\caption{Hole in the horizon of a generic rotating black hole
formed by a singular beam directed along axis
of symmetry.}
\end{figure}

The two event horizons are joined into one connected surface, and
the surface of the event horizon lies inside the boundary of the
ergosphere. As a consequence, the axial singularities lead to the
formation of holes in the BH horizon, thus opening the interior of
the BH to external space.

\section{Aligned wave excitations and beam-like pulses}

 Corresponding exact solutions with wave
electromagnetic excitations {\it on the Kerr background} have been
obtained and investigated in \cite{BEHM,BurAxi}. They are
asymptotically exact in the low-frequency limit, i.e. are also
self-consistent, taking into account back reaction. In the far
zone the wave beams were considered in \cite{BurAxi} and it was
shown that locally they have the structure of the self-consistent
exact $pp-$wave (Peres) solutions.
Note, that similar singular `axial' strings where obtained long
ago in the Robinson-Trautman class of solutions \cite{KraSte}, and
also for fermionic wave excitations on the Kerr geometry
\cite{EinFin}.

Wave excitations propagating in the direction $Y_i $  are
described by means of the function \be \psi_i(Y,\t)
=q_i(\t)\exp\{i\omega \t\}\frac 1 {Y-Y_i}, \label{E8}\ee where
$q_i(\t)$ is a slowly varying amplitude of excitation with
analytic dependence on the retarded time $\t.$ Similar to the
usual treatment of the plane waves, such infinitely extended
beam solutions have to be replaced by physically realizable
solutions with finite extension. In particular, the wave and
nonstationary generalizations of these solutions have to contain
some time-dependent amplitude factor $q=q(\t)$ which turns them
into beam-like pulses. The pp-wave singularity can also be
regularized, as it was considered in \cite{BurAxi}.

The equations for the gravitational sector, obtained by Debney,
Kerr and Schild (DKS), are given in App.~A. The most
essential fact in the treatment of these wave and non-stationary
 Kerr-Schild solutions is the appearance of an extra radiation
field $\gamma,$ which is absent in the term $H$ (\ref{Hpsi}) and
does not have immediate influence on the metric tensor. However,
it leads to the appearance of  null radiation $\cF_\mn =\gamma Z
e^3_\m e^1_\n $ which is described by the Vaidya shining star
solution \cite{KraSte}. This radiation leads to the loss of mass
of the black hole, and we show in App.~A that it is small for
low-frequency solutions, since the rhs of the gravitational
equations $\gamma \sim \dot \psi$ tends to zero in the
low-frequency limit $\omega \rightarrow 0 .$ Therefore, the wave
beam-like pulses  tend to self-consistent Kerr-Schild solutions in
this limit.

The structure of the solutions contains the following very
important physical peculiarity: the smallness of the function $\gamma$
does not mean that the function $\psi(Y,\t)$ in (\ref{Hpsi}) is
required to be small too. The electromagnetic field determined by
$A=\psi/P^2$ and the corresponding distortion of the metric and
horizon by the function $H$ are independent from the frequency and
can be in fact very strong.
Note also, that there is one subtle point in our treatment related
with the fact that, due to the poles in the angular directions
$Y_i$, the limit $\gamma\to 0 $ as $\omega \to 0 $ is not a
uniform one, which demands a thorough treatment. However, the general
integral expressions for $\gamma$ contain a free term, and in
App.~B we show that this term may be used for the regularization
of these poles, without any influence on the function $A$.

\section{Aligned wave excitations and micro-holes in the horizon}

Since the aligned solutions here considered tend to the exact ones
in the asymptotic limit $\gamma \to 0$, this also means,
physically, that they must tend to the exact solution
corresponding to `ringing modes' of a black hole subjected to a
weak external electromagnetic excitation. In particular, it may be
the vacuum field of virtual photons. It could be reduced to a
standard quantum treatment of scattering plane waves on a
potential barrier, if only the extra Kerr-Schild demand of the
alignment of the exact solutions with the Kerr congruence
(\ref{align}) would be obeyed. However, this demand is very hard,
and the plane waves cannot reside in the Kerr background. Even in
the limit of zero mass term, the Kerr background is not Minkowski
one and has a two-sheeted topology which does not admit plane waves.
This situation is close related with the twistor analytic structure
of the Kerr geometry, and prevents the use of a standard scattering
approach. The character of this paper does not allow us to discuss
this problem. \fn{The replacement of plane waves by twistor null
lines (beams) was shown by Witten in the new approach to
scattering amplitudes in twistor space \cite{Wit}, inspired by
pioneering work of Nair \cite{Nai}.} Because of that, we will follow
the simple idea of York \cite{Yor} to consider the back-reaction
of an elementary excitation of a black-hole caused by an
elementary interaction with a virtual photon. In most of the
previous treatments of this sort, deformation of the horizon was
related with the replacements $m \to m +\delta m$ and $J\to
J+\delta J$, retaining the spherical topology of the horizon. In
our case, in accordance with the discussed solutions, an
elementary excitation aligned with the Kerr congruence represents
a light-like beam, which in the far zone takes the form of
$pp$-wave string-like solution, and the corresponding elementary
deformation of the horizon changes its topology, even at the
low-frequency limit.

We neglect the recoil and use expressions (\ref{PY3}) for the
stationary case, which allow us to get the exact solution for
function $A$ in the form containing the wave excitations
\be A=\psi(Y,\t)/P^2  . \label{Atot1}\ee
This equation  is linear in $\psi$  and the total excitation
caused by the virtual photons will be a sum over elementary
excitations in the distinct directions $Y_i ,$
\be \psi(Y,\t) = \sum_i \frac {q_i(\t)} {Y-Y_i} \exp\{i\omega_i
\t\}. \label{psiYt} \ee
The wave solution (\ref{psiYt}) with many excitations will become
exact on the Kerr background, while the back reaction will break
 self-consistency, leading to some disclosure in the
gravitational sector, as it is discussed in App.~B, this
disclosure is proportional to the term $A \bar\gamma$ in the
equations (\ref{G7}), and has to tend to zero in the low-frequency
limit together with $\gamma \to 0$.  The corresponding retarded
time has the form \cite{BurAxi} $ \t = t -r -ia \cos \theta $,
which allows us to obtain the general retarded-time solution $
\gamma =\gamma_0 +\gamma_f \label{gam1}$
 as the sum of the partial solution $\gamma_0 $  containing series of poles,
\be \gamma_0 = \sum _i c_i(\t) \frac 1{P^2Y(Y-Y_i)} \ee with
oscillating factors
\be c_i(\t)=i \omega 2^{1/2}q_i(\t)\exp\{i\omega_i \t\}
\label{ci1} , \ee and the term
\be \gamma_f = \frac{\phi (Y,\t)} {P} \label{gamf1} \ee which is
determined by a free function $\phi (Y,\t).$

The free term $\gamma_f$ has to be tuned to have the same series
of poles and the same oscillating factors of opposite sign to
provide regularization of poles by subtraction. It is shown in
App.~B that such a regularization does not touch the function
$A$ and provides a uniform low-frequency limit $\gamma_{out} \to 0$.
Following to analogues with family of Green functions one can
introduce the step function $\theta(r)= \pm 1 ,$ separating the
solutions on the positive sheet $r>0$ from the solutions on the negative one,
$r<0 ,$ and consider the ingoing field of virtual photons as a
non-regularized one, $\gamma_{in} =\gamma_0 ,$ while the field $A$
and the regularized field $\gamma_{reg}=\gamma_{in} -\gamma_{comp} $
may be considered as outgoing radiation $\gamma_{out}=\gamma_{reg} .$

The question can be asked, what is the mechanism which provides
such a compensating regularization and discontinuity of the
solutions? We do not know a precise answer, although it is quite
clear that the discontinuity has to be related to some source\fn{The
problem of the Kerr source is one of the most hard and ambiguous.
There are evidences of a stringy structure of the Kerr source
\cite{BurOri,BurSen,BurTwi}.} positioned in the vicinity of the
Kerr disk $r=0 ,$ and that regularization minimizes the interaction
between the vacuum field $\gamma_{in}$ and the one induced by the Kerr
source excitations, in the form of the field $A$. The process of
scattering may be interpreted in this case as a type of scattering
of the incoming vacuum radiation on the stringy structure of the
Kerr geometry.

It should be emphasized  once more that there is a principal
difference between the roles played by the fields $A$ and $\gamma
.$ The fields $\gamma$ as well as $\gamma_{reg}$ do not act
immediately on the metric, being respectable ones for the radiation and
the appearance of singular micro-beams covering the celestial sphere
$S^2.$. Its influence on the metric is indirect and occurs only
via the loss of mass by radiation, in accordance with
expression (\ref{G8}), where  $\gamma_{reg}$ has to be
substituted by the outgoing radiation, $\gamma_{out}=\gamma_{reg}.$
At the same time, the exact solutions for the electromagnetic field
$A$ on the Kerr-Schild background, have direct influence on the
metric, leading to the formation of the holes at the horizon.  The direct
influence of the field $A$ on the metric and on the horizon will
result into formation of multiple micro-holes in the horizon
and its topological fluctuations, as depicted in Fig.~2.

\begin{figure}[ht]
\centerline{\epsfig{figure=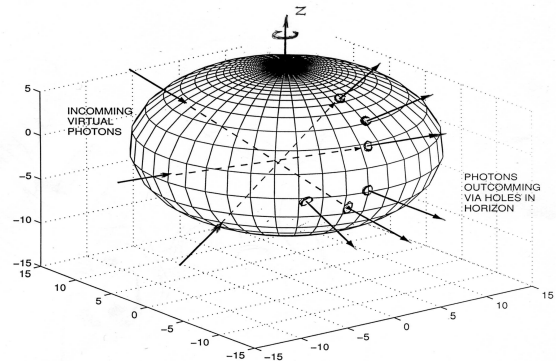,height=6cm,width=8cm}} \vspace*{-7mm}

\caption{Excitation of a black hole by the zero-point field of
virtual photons forming a set of micro-holes at its horizon.}
\end{figure}

In the far zone the stress tensor of the electromagnetic field is
determined by the leading term $\cF _{31} ,$ (\ref{E6}), which
describes the null electromagnetic radiation along the rays of the
congruence $e^3_\m=P k_\m .$ At the same time, the stress-energy
tensor  is represented by the $T_{33}$ component, which is determined
by the $\gamma-$term \be T_{\m\n} = T_{33}e^3_\m e^3_\n = \frac 1 2
\gamma_{out}\bar \gamma_{out} P^2 k_\m k_\n .\ee Radiation
propagates along the Kerr congruence which is geodesic with the
null tangent vector $k^\m$. This means that radiation is massless.
Each elementary i-th excitation, with a pole $\sim \frac 1 {(Y-Y_i)}$,
propagates along the i-th null ray in the direction $k^\m(Y_i).$
The in-going radiation propagates from $r= -\infty$ on the
negative sheet of the metric towards the Kerr disc $ r=0.$ The
outgoing radiation propagates on the positive sheet in the
direction $r \to \infty .$ The averaged term on the rhs of the
gravitational equation (\ref{G8}), $<\gamma_{out}\bar\gamma_{out}>
= \sum_i <\gamma_{i out}\bar\gamma_{i out}> $ is positive and
determines the incoherent radiation corresponding to the known
Vaidya shining-star solution \cite{Yor,KraSte}.

\section{Discussion}

In a number of recent works, the origin of thermal
emission from black holes has been related with conformal-analytical
structures inherent to two-dimensional, 1+1 dimensional and 2+1
dimensional black holes
\cite{Strom,Carl,Carl1,CousHen,Banad,Card}.
The two-dimensional case is the most familiar and the
corresponding conformal field theories have been elaborated in detail,
including corresponding quantum gravity issues. Quantum effects of the low
dimensional black holes are related to some Virasoro algebra
which, on the other side, represents the core of the (super)string
theory. Therefore, one observes the growing tendency that the methods of
superstring theory acquire increasing meaning in the physics of
black holes. The Kerr-Schild structure of the Kerr geometry is
based on the Kerr theorem, the complex-analytic structure of which
is expressed in terms of twistors. Twistorial analyticity of the Kerr
geometry \cite{BurTwi} is a {\it natural} four-dimensional analog
to two-dimensional conformal-analytic stringy structures. The
analyticity of electromagnetic excitations of the Kerr geometry is
expressed in terms of the alignment of the electromagnetic fields
with the holomorphic structure of the Kerr congruence, which is
determined by the Kerr theorem in terms of an analytic surface in
twistor space \cite{BurTwi,Multiks}.

A central role in the twistorial structure is played by the
 two-dimensional sphere
$S^2, $ parametrized by the analytic function $Y(x)$ which is a
projective angular coordinate. The singular Kerr-Schild solutions
are described by holomorphic fields on this sphere, just similar
to topological excitations of the WZW model.
Therefore, the mechanism of the origin of the quantum spectrum in
the Kerr-Schild geometry is to be  similar to the suggested one for
2- and (2+1)-dimensional BTZ black holes, which is also
based on the topological WZW model \cite{Carl1,Banad}. In
particular, it was obtained in \cite{Kim} that a BTZ black hole
may be represented as a section of the four-dimensional Kerr black
hole in the equatorial plane $\theta=\pi /2 .$\fn{At first sight
this looks strange, since a BTZ black hole has a negative scalar
curvature, contrary to the zero value it has in the Kerr solution. However,
there is a very simple argument in favor of this statement.
Indeed, separating the $\theta$ variable in the Kerr solution, one
has to remember that it is an angular variable giving a positive
contribution to the total scalar curvature of the Kerr space-time,
which is zero. Consequently, the scalar curvature $R_{2+1}$ of the
2+1 factor-space will be negative, in full correspondence with the BTZ
black hole.} It is known that the dual string models \cite{GSW} are based on
 conformal field theory \cite{BPZ,KniZam,GSW}, the core of which is formed by quantum
oscillators obeying a Virasoro algebra \cite{GSW} \bea [L_m, L_n ]
&=& (m-n) L_{m+n} + \frac c {12} (m^3 -m) \delta_{m+n,0},
\label{Vir} \eea with similar expression for conjugate generators
$\bar L_n .$ The central charge  $c$ depends on the details of the
quantum system considered and was studied for a large variety of
different stringy models \cite{GSW}. However, numerous
investigations of different models have shown \cite{Carl1} that
details of the theories are not important, and that the key requirement
is the existence of the conformal structure, in particular, the $S^1$
isometry providing the algebra of the gauge field transformations
Diff $S^1 .$ The Bekenstein-Hawking thermal spectrum is reached in
most of these models on the base of Cardy's formula \cite{Card},
with independence of the value of the central charge
\cite{Strom,Carl1}.
Similarly, the conformal group of the Kerr-Schild geometry forming
the core of its analytic properties is  related with the
projective angular coordinate $Y(x),$ which is one of the three
twistor coordinates of the projective twistor space. It has the
stringy structure \cite{BurStr,BurDir} and may determine the spectrum of the radiation from the
Kerr black hole similar to BTZ black holes in accordance with Cardy's
formula. This demands extra considerations.

The main goal of this paper was to show the existence of a
 very nontrivial beam-like
form of excitations and of a related radiation of the Kerr-Schild
geometry, as well as a quite nontrivial form of a corresponding
topological deformation of the horizon, which are based on analytic
properties of the Kerr-Schild geometry and determined by the
alignment conditions for the electromagnetic excitations.
In conclusion, we  touched upon the old question: where is
the black hole radiation created? Contrary to the familiar point
of view that radiation is created near the horizon, we arrive at
the conclusion that the horizon plays a rather passive role. Being
perforated by a series of beams, it provides a
way to escape for radiation, while the creation of the radiation itself is related
with the gravitational anomaly which occurs under regularization of
the poles of the energy-momentum tensor. In this respect we share
the point of view discussed in \cite{RobWil}. Indeed, in the case
of a stationary field $A,$ i.e. $\dot A =0$, the field $\gamma$ is
decoupled, the free flow of zero-point radiation $
T_{(\gamma)}^{\m\n} = \frac 1 2 \gamma\bar \gamma P^2 k^\m k^\n $
obeys the conservation law $\d _\m T_{(\gamma)}^{\mn} =0$, and the
field $\gamma$ may be decoupled from this system of equations
\cite{BurAxi,BurOri}. Creation of the wave field $A$ is similar to
a scattering process which, in accordance with \cite{RobWil},
leads to an anomaly in general covariance, taking the form of a
non-conservation of the energy-momentum tensor and signaling towards
the interaction with some source in the core of the black hole, which
is apparently the true source of radiation.

This conjecture demands extra investigations of the stringy Kerr
source at the quantum level, which will be the object of further
work and is beyond the scope of this paper.

\section*{Acknowledgement}

We thank Valery Frolov and Theo Nieuwenhuizen for useful
discussions on the earlier stage of this work. This work was
supported by grant RFBR 07-08-00234, by MEC (Spain),
BFM2006-02842, PR2006-0145, and by AGAUR, grant 2007BE-100003 and
contract 2005SGR-00790.
\medskip

{\bf Appendix A.} We consider here the equations for the
gravitational sector obtained by Debney, Kerr and Schild (DKS)
 \be M,_2 - 3 Z^{-1} \cZ Y,_3 M = A\bar\gamma \cZ , \label{5}\ee
\be \cD M = \frac 12 \gamma\bar\gamma  , \quad M,_4=0 \label{6}\ee
where $\cD$ is given by \be \cD=\d _3 - Z^{-1} Y,_3 \d_1 - \cZ
^{-1} \Y ,_3 \d_2   \
 \label{cD} .\ee
We assume that the mass of the black hole $m$ is much bigger than the
energy of the excitations, neglect the possible recoil,
and use the expression for the stationary function $Y(x)$ and its
tetrad derivatives. This yields
 \be P=2^{-1/2}(1+Y\Y), \quad Y,_3= - \bar Z P_\Y /P , \label{PY3} \ee
and allows one to reduce the equations (\ref{5}) and (\ref{6}) to
the very simple form \be m,_\Y = A\bar \gamma P^3 \ , \quad A =
\sum_i \psi_i / P^2  \ , \label{G7} \ee \be \dot m =-\frac 12 P^4
\gamma \bar \gamma \ .\label{G8}\ee It is known \cite{KraSte} that
the last equation determines the loss of mass by radiation. The
right sides of both gravitational equations will be small for the
low-frequency aligned wave excitations, since the function
$\gamma$ will be of the order $\gamma\sim \dot\psi \sim i\omega
\psi .$
\medskip

{\bf Appendix B .} We consider here in more detail the
Kerr-Schild  equations obtained in \cite{DKS} for the
electromagnetic sector
 \be A,_2 - 2 Z^{-1} \cZ Y,_3 A = 0 , \quad A,_4=0 , \label{3}\ee
\be \cD A+ \cZ ^{-1} \gamma ,_2 - Z^{-1} Y,_3 \gamma =0, \quad
\gamma,_4=0 , \label{4}\ee where $\cD$ is given by (\ref{cD}). We
neglect the recoil and use expressions (\ref{PY3}) for the
stationary case, what allows us to get the exact solution for the
function $A$ in a form containing the wave excitations \be
A=\psi(Y,\t)/P^2  , \label{Atot}\ee and reduce Eq.~(\ref{4}) for
the adjoined field $\gamma$ to the very simple form \be \dot A=
-(\gamma P),_\Y . \label{11} \ee These equations  are linear in
$\psi$ and $\gamma ,$ and the total excitation caused by the
virtual photons will be a sum over elementary excitations in the
distinct directions $Y_i ,$ $ \psi(Y,\t) = \sum_i \frac {q_i(\t)}
{Y-Y_i} \exp\{i\omega_i \t\}. $

The wave solution (\ref{psiYt}) with many excitations will be {\it
exact on the Kerr background}, while the back reaction will break
 self-consistency, leading to some disclosure in the
gravitational sector. As it was discussed in the previous section,
this disclosure is proportional to the term $A \bar\gamma$ in
Eqs.~(\ref{G7}), and has to tend to zero in the low-frequency
limit together with $\gamma \to 0$. However, the limit $\gamma\to
0 $ will not be uniform because of the poles, and a thorough
procedure is necessary for the `regularization' of singularities.

The equation (\ref{11}) can be integrated in the general form
\cite{BurAxi}. To perform it, one should introduce the
retarded-time parameter $\t$ obeying the constraints\fn{These
condition means that the gradient $d \t$ is spanned by the null
vectors $e^1$ and $e^3,$ and that the complex planes $\t=$const.~are
null and tangent to the light cones. Note also, that
these constraints extend to all the projective twistor coordinates
$Z^A$  \cite{BurTwi}.} satisfy $ \t,_2=\t,_4=0 . $ The corresponding
retarded time has the form \cite{BurAxi} $ \t = t -r -ia \cos
\theta ,$ which allows us to obtain the general retarded-time
solution $\gamma =\gamma_0 +\gamma_f \label{gam}$
 as the sum of the partial solution $\gamma_0 $
  containing series of poles,
$ \gamma_0 = \sum _i c_i(\t) \frac 1{P^2Y(Y-Y_i)} $ with
oscillating factors $ c_i(\t)=i \omega
2^{1/2}q_i(\t)\exp\{i\omega_i \t\} \label{ci} , $ and the term $
\gamma_f = \frac{\phi (Y,\t)} {P} \label{gamf} $ which is
determined by a free function $\phi (Y,\t).$

The free term $\gamma_f$ could be taken with the same series of
poles and the same oscillating factors of opposite sign to provide
regularization of poles by subtraction. However, the extra slowly
varying function $P=2^{-1/2}(1+\Y Y)$ has the different degrees in
the front of the functions $\phi$ and $\psi ,$  which prevents
immediate compensation. Working in a complex vicinity of i-th
pole, one can set $\Y = \Y_i$ and expand the function $P(\Y_i,Y),$
in $(Y - Y_i) ,$
\be P(\Y_i, Y) =P_i + 2^{-1/2} \Y_i (Y-Y_i) + {\cal O} [(Y-Y_i)^2]
, \label{Pan} \ee where $P_i = 2^{-1/2}[1+\Y_i Y_i] .$
Then the free function $\phi (Y,\t)$ chosen in the form $ \phi
(Y,\t)= - \sum_i   c_i(\t) \frac 1 {Y_i P(\Y_i,Y) (Y-Y_i)}, $ will
regularize the solution, compensating the poles in $\gamma_0 .$  The
result of this compensation is the function
\be \gamma_{reg}=  - \sum_i c_i(\t) \frac {2^{-1/2} \Y_i + {\cal
O} (Y-Y_i) }{Y_i P P(\Y_i,Y)}.   \ee Such a regularization does
not touch the function $A$ and provides a uniform low-frequency
limit $\gamma_{out} \to 0 .$ The regularized function
$\gamma_{out}$ represents a stochastic process which is the sum of
many oscillating terms. The mean value of this process $<\gamma>$
is zero. Looking on the product $A\bar\gamma ,$ one sees that the i-th
pole in function $A$ is compensated by the k-th zero $(Y-Y_k)$ in
$\bar \gamma$ by setting $i=k , $ while for $i\ne k ,$ the product averaged
over time $A\bar\gamma$ vanishes due to the difference in the
frequencies of oscillations. Therefore, we pick up the exact
wave solutions of the electromagnetic sector, for which the correlation
$<A\bar\gamma
>$ contains a sum of regular terms tending to zero in the
low-frequency limit. Similar treatment of the correlation $<\bar
\gamma_{out} \gamma_{out}>$ shows that the terms with $i\ne k$ do
not correlate and vanish, while the sum with $i=k$ survives and
turns out to be singular, displaying radiation along the Kerr
congruence, which is similar to the radiation of the Vaidya shining
star solution.

\end{document}